\documentclass[runningheads]{llncs}
\usepackage{amssymb}
\usepackage{amsmath}
\usepackage[english]{babel}
\usepackage{graphicx}

\begin{document}
\title{LDU factorization}
\titlerunning{LDU factorization}
\author {Gennadi Malaschonok \thanks{The work was partially supported by the Academy of Sciences of Ukraine,  project (2020) "Creation of an open data e-platform for the collective use centers".} }
\authorrunning{Gennadi Malaschonok }
\institute{  National University of Kyiv-Mohyla Academy, \\
2 Skovorody vul., Kyiv 04070, Ukraine \\
\email{malaschonok@gmail.com }}

\maketitle

\begin{abstract}
 LU-factorization of matrices is one of the fundamental algorithms of linear algebra. The widespread use of supercomputers with distributed memory requires a review of traditional algorithms, which were based on the common memory of a computer. Matrix block recursive algorithms are a class of algorithms that provide coarse-grained parallelization. The block recursive LU factorization algorithm was obtained in 2010. This algorithm is called LEU-factorization. It, like the traditional LU-algorithm, is designed for matrices over number fields. However, it does not solve the problem of numerical instability. We propose a generalization of the LEU algorithm to the case of a commutative domain and its field of quotients. This LDU factorization algorithm decomposes the matrix over the commutative domain into a product of three matrices, in which the matrices L and U belong to the commutative domain, and the elements of the weighted truncated permutation matrix D are the elements inverse to the product of some pair of minors. All elements are calculated without errors, so the problem of instability does not arise.
\end{abstract}

\section*{Introduction}

The representation of the matrix $ A $ in the form of two factors $ A = LU $, where L is the lower triangular matrix and U is the upper triangular matrix, is called the LU decomposition (or LU factorization). This decomposition is considered as the basic algorithm for any library of linear algebra programs \cite{1}. Many algorithms are built on its basis, including solving linear systems, matrix inversion, rank calculation, etc.

With the advent of supercomputers, it became possible to increase the size of matrices in solving applied problems. At the same time, shortcomings of the known matrix algorithms appeared. It became clear that they cannot be applied to large matrices. Problems such as accumulation of rounding errors, loss of accuracy, poor concurrency, loss of sparseness of matrices, high computational complexity, and other problems began to appear (see, for example, \cite {0}).

In 2010, the LEU factorization algorithm was obtained. This is an algorithm that applies to matrices over number fields. It allowed to overcome many of these shortcomings for finite number fields. It was the first block recursive algorithm with the complexity of matrix multiplication, sparse and highly parallelistic \cite {2}.  

However, the problem of loss of accuracy, in the case of classical fields, remained as before impossible to overcome. It was required to obtain a generalization of LEU factorization to commutative domains and their fields of quotients. Such an algorithm will allow, for example, to use integers in the calculations instead of approximate rational numbers. Two approaches were proposed to create such an LDU factorization algorithm \cite {4}, \cite {5}. The present work continues and completes these studies. We propose the complete dichotomous recursive LDU factorization algorithm for the commutative domain and give its proof.
\section{Initial statement of the problem}
 Let $R$ be a commutative domain, $F$ its field of quotients. Let $A\in R^{n\times n}$  be a matrix that has rank $r$, $ r\le n$. We want to get matrices $L,U \in R^{n\times n}$ of rank $n$,( L is  low triangular, U is upper triangular), matrix $D $, that has rank $r$,  with $r$ non-zero elements equal $(\ det_{1})^{-1}$, $(det_{1} det_{2})^{-1}$, .., $(det_{r-1} det_r)^{-1}$, such that 
  $$ A=LDU. $$
  We denote by $ det_r $  the $ r \times r $ nonzero minor of the matrix $ A $, whose position is determined by $r$ nonzero rows and columns of the matrix $ D $.
  The determinants of successively nested nondegenerate submatrices of orders $ r-1, r-2, .., $ 2.1 we denote $ det_ {r-1} $, $ det_ {r-2} $, .., $ det_ { 2} $, $ det_ {1} $, respectively.

To solve this problem, we formulate a more general problem, but first give some necessary definitions.
 \section{ Preliminary information}
 \subsection{ Semigroup of truncated weighted permutations}
   The diagram shows the structure of the semigroup of truncated weighted permutations ${\bf S}_{wp}$:
  
  \begin{figure} [h]
\begin{center}
\includegraphics[scale=0.5]{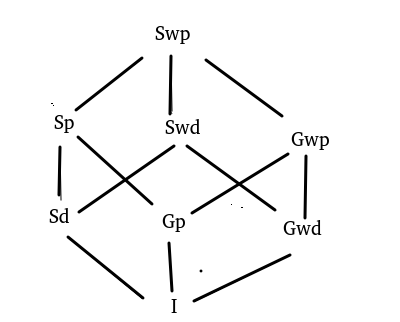}
  \caption{The structure of the truncated weighted permutations semigroup ${\bf S}_{wp}$}
 \label{graphinv}
 \end{center}
\end{figure}
  At the center of the diagram you can see a permutation group $ {\bf G} _p $.
  Between $ {\bf S}_{wp} $ and $ {\bf G}_p $ there are two more subalgebras: 
  ${\bf S}_p$ and ${\bf G}_{wp}$. 
 
 If we replace elements with values of 1 in the matrices from the permutation group $ {\bf G}_p $ by arbitrary nonzero elements, then we obtain the group of weighted permutations ${\bf G}_{wp}$.
 
 If, on the contrary, in the matrices from the permutation group $ {\bf G}_p $ we replace some elements with values 1 with zero elements, we obtain a semigroup of truncated permutations $ {\bf S}_p $. 
 
The semigroup of truncated weighted permutations $ {\bf S}_ {wp} $ is obtained from the group $ {\bf G}_{wp} $ if we replace some nonzero elements with zeros.

If we select all diagonal matrices in the semigroup $ {\bf S}_{wp} $, then we obtain a semigroup of (weighted) diagonal matrices 
$ {\bf S}_{wd} $. Two other subalgebras $ {\bf S}_d $ and $ {\bf G}_{wd} $ are embedded in it.
The semigroup $ {\bf S}_d $ is formed by diagonal matrices for which only 1 and 0 can stand on a diagonal. The group $ {\bf G}_{wd} $ is formed by those diagonal matrices from $ {\bf S}_{wp } $ for which there are no zero elements on the diagonal. The identity group $ \bf I $ that contains one identity matrix closes this construction.
 \subsection{Some mappings on semigroups}
  For matrices from the semigroup $ {\bf S}_{wp} $ we introduce two mappings:    {\bf unit} and {\bf extended}.

  A homomorphism of the multiplicative groups $ F^* \to 1 $ (or $ R^* \to 1 $) induces a homomorphism of the corresponding subalgebras:
  ${\bf S}_{wp} \to {\bf S}_{p}$, ${\bf S}_{wd} \to {\bf S}_{{\widehat D}}$, ${\bf G}_{wp} \to {\bf G}_{p}$, ${\bf G}_{wd} \to I$. 
All nonzero elements of the matrix are replaced by unit elements. On the diagram, they correspond to arrows that are directed down and to the left.
   \begin{definition} 
   The mapping of the matrix $ D \in {\bf S}_{wp} (F) $ induced by the homomorphism 
   $F^* \to 1$:
 $$  D\in {\bf S}_{wp} \to  D^{\to 1} \in {\bf S}_{p}$$
 is called   {unit mapping}. 
  \end{definition}
The unit homomorphism of semigroups can be represented by the following commutative diagram:


 $\hspace{40mm}    (A,B) \xrightarrow{\ \ \to 1 \ \ } (  A^{\to 1}, B^{\to 1})$

 $\hspace{40mm}   \  \times \downarrow  \ \ \ \ \ \ \ \ \ \times \downarrow  $

 $\hspace{40mm}  \ \ \   C \ \xrightarrow{\ \ \to 1 \ \ } \   C^{\to 1}$

\begin{definition}
The matrix mapping $D\in {\bf S}_{wp} \to   D^{Ext} \in {\bf G}_{wp} $ in which every block at the intersection of zero rows and zero columns is replaced by a unit block called {\bf extended mapping} and is indicated by a "Ext" upper index.
\end{definition}
 On the diagram, the extended mapping corresponds to 4 arrows that are directed down and to the right:   ${\bf S}_{wp} \to {\bf G}_{wp}$, ${\bf S}_{wd} \to {\bf G}_{wd}$, ${\bf S}_{p} \to {\bf G}_{p}$, ${\bf S}_{{\widehat D}} \to I$.  
 As a result of such a transformation, a matrix of full rank is obtained.
 
 \begin{definition}
    The mapping of the matrix $D\in {\bf S}_{wp} \to  \bar  D =D^{Ext}-D \in {\bf S}_{p}$ 
    is called the {\bf  complementary mapping} and is denoted by a ``bar''.
 \end{definition}

\begin{property}
The special case of the complementary mapping, when the matrix $D$ belongs 
to the semigroup ${\bf S}_p$, is an involution on the semigroup ${\bf S}_p$. 
Involution is reversible: $\bar {\bar D}=D$.
\end{property}%
\begin{property} $ \forall D \in {\bf S}_p: D+\bar D \in {\bf G}_p $.
\end{property}%
Examples of such involution:  $\left(\begin{array}{ccc}0  & 0 & 0\\ 0  & 0 & 0\\ 1  & 0 & 0 \end{array}\right) 
\to  \left(\begin{array}{ccc}0  & 1 & 0\\ 0  & 0 & 1\\ 0  & 0 & 0  \end{array}\right)$,
$\left(\begin{array}{ccc}0  & 0 & 0\\ 0  & 1 & 0\\ 0  & 0 & 0 \end{array}\right) 
\to  \left(\begin{array}{ccc}1  & 0 & 0\\ 0  & 0 & 0\\ 0  & 0 & 1  \end{array}\right)$.

\subsection {Surrounding  minors}
  Hereinafter, we consider matrices over the commutative domain $ R $.
 \begin{definition}
Let a matrix $ \cal M $ be given and let $ A $ be its square submatrix located in the upper left corner. Any submatrix $ G$, that is obtained by adding to the block $ A $ some row $ r $ and some column $ c $ of matrix $ \cal M $ 
   $$ G= \left(\begin{array}{cc}A  & c \\ r & \omega \end{array}\right).  \ \ \eqno(1) $$
is called the submatrix that surrounds matrix $ A $.
 \end{definition}
 \begin{theorem}
 Let $ A $ be a square matrix, $ A^* $ the adjoint matrix for $ A $, $ det (A) \ne 0 $, $ G $ the surrounding  matrix for $A$ (1), then 
 $$det(G)=  det(A) \omega - r A^* c.$$
 \end{theorem}
 {\bf Proof.}
 This equality expresses the decomposition of the determinant $ G $ in row $ r $ and column $ c $.
 %
 %
\begin{theorem} 
Let a matrix $ \cal M $ be divided into blocks
      $ {\cal M} = \left (\begin {array} {cc} A_k & B \\ C & D \end {array} \right), $
     $ A_k $ is a square block of size $ k \times k $, its determinant $ det_k $ is non-zero and $ A_k^* $ is an adjoint matrix for $ A_k $, then the elements of the matrix
     $$ A = (det_k) D - C A_k ^ * B $$
     are the minors that surround the $ A_k $ block.
 \end{theorem}
 {\bf Proof.}
 To prove the theorem, it suffices to apply Theorem 1 to each element of the matrix $A$.
 %
 %

 \section{Statement of the problem} 
 \subsection{General statement of the problem} 
  Let $R$ be a commutative domain, $F$ its field of quotients.
  Let a matrix $ {\cal M} \in R^{ N \times N} $ ($N=2^\nu$) be given, and let $ \alpha $ be the determinant of the largest nondegenerate (or empty) submatrix located in the upper left corner of the matrix $ {\cal M} $. For the case of an empty submatrix, we set $\alpha = 1 $.
  
  Let $ n = 2^p $, $ A \in R^{n \times n} $ be a matrix of rank 
  $ r $, $ r \le n $, and elements of the matrix $ A $ be surrounding  minors with respect to the minor $ \alpha $. In the case of an empty submatrix, we can take $ A = \cal M $.
  
  We want to obtain matrices $ L, U, M, W \in R^{n \times n} $ of rank $ n $, (L lower triangular, U - upper triangular), a   matrix $D \in {\bf S}_{wp}(F)$, of rank $ r $,
   with $r$ non-zero elements equal $(\alpha \det_{1})^{-1}$, $(det_{1} det_{2})^{-1}$, .., $(det_{r-1} det_r)^{-1}$,
  such that
 $$
 \left\{ \begin{array}{rcl} \alpha L D U = A  \\  L \widehat D M =  {\bf I } \\  W \widehat D U =   {\bf I } \end{array}\right. .
 \eqno(1)
 $$
  We denote by $ det_r $  the $ r \times r $ nonzero minor of the matrix $ A $, whose position is determined by $r$ nonzero rows and columns of the matrix $ D $.
  The determinants of successively nested nondegenerate submatrices of orders $r-1, r-2,.., 2, 1$   we denote $ det_ {r-1} $, $ det_ {r-2} $, .., $ det_ { 2} $, $ det_ {1} $, respectively. 
  
  We denote  by $ \widehat D $ a matrix 
  $$\widehat D = det_ {r}^{-1}(\alpha D)^{Ext}=\alpha det_ {r}^{-1}  D  + 
  det_ {r}^{-1} \bar D,  $$   
   and we denote  
  $$  E =D^{\to 1}, I=EE^T, J=E^TE, \bar I={\bf I }-I, \bar J={\bf I }-J,\eqno(2)$$
  $E  \in {\bf S}_{p},\ I,J\in \bf  S_{d}$, ${\bf I }$ is the unit matrix.  
  
  It should be noted that the matrices $ I $ and $ D $ have the same nonzero rows, and the matrices $ J $ and $ D $ have the same nonzero columns: 
$$ I D J = D,\  \ \bar I D  =0,\ \ D \bar J = 0.  $$
We define the properties of the matrices $ L $ and $ U $ as follows:
 $$L\bar I=\bar I ,\ \ \    \bar J U=\bar J.  \eqno(3)$$ 
%

\section{ Dichotomous Recursive Decomposition Design} 
 We want to describe a procedure that allows you to compute the LDU-factorization 
 $$
 (L, D, U, M, \widehat D, W, \alpha_r ) = {\bf LDU}(A, \alpha)
 $$
 in a recursive and dichotomous way.

 (1)   If $A=0$, then we assume that  $D=I=J=0$, $M=W=\alpha \bf I$,  ${\widehat D}=\alpha^{-1} \bf I$,
 $L=U=\bar I=\bar J=\bar D=\bf I$, $\alpha_r=\alpha$.

\noindent 
 (2) If $n=1$ ($A=[a], a \ne 0$), then  we assume that  $D=[(\alpha a)^{-1}]$, ${\widehat D}=[ a^{-2}]$, $L=U=M=W=[a]$, $ I= J= [1]$, $\bar I=\bar J=\bar D=[0]$, $\alpha_r=  a$.
 
\noindent
 (3) For $A\ne 0$ and $n>1$ we divide matrix $A$ into four equal blocks 
 $$
 A= \left(\begin{array}{cc}A_{11} & A_{12} \\ A_{21} & A_{22} \end{array}\right).  \eqno(4)
 $$
We can do the LDU-decomposition for block $A_{11}$:
 $$
 \left\{ \begin{array}{rcl} \alpha L_{11}D_{11}U_{11}= \ A_{11} \\  L_{11}{\widehat D}_{11}M_{11} =   I \\  W_{11}{\widehat D}_{11} U_{11}=I  \end{array}\right.. \eqno(5)
 $$ 
 Let $k=\, \mathbf{rank}(A_{11})$ and $det_{1}, det_{2}, .., det_k $ are the non-zero nested leading minors of $A_{11}$, that were found recursively, $\alpha_{k}=det_k$,   
 ${\widehat D}_{11}= \alpha_k^{-1} (\alpha D_{11}+ \bar D_{11})$
 and the non-zero elements of $D_{11}$ equal $(\alpha det_{1})^{-1}$, $(det_{1} det_{2})^{-1}$ ... $(det_{k-1} \alpha_k)^{-1}$.

 Then we can write the equality 
  $$    \left(\begin{array}{cc}A_{11} & A_{12} \\ A_{21} & A_{22} \end{array}\right)  = \left(\begin{array}{cc}L_{11} & 0 \\ 0 & I \end{array}\right)   \left(\begin{array}{cc}  \alpha D_{11} & { 1 \over \alpha_{k}} A'_{12} \\  { 1 \over \alpha_{k}} A'_{21} &    A_{22} \end{array}\right)   \left(\begin{array}{cc}U_{11} & 0 \\ 0 & I \end{array}\right) . \eqno(6)$$ 

 We denote the new blocks:

 $$
 A'_{12}=   { \alpha_{k} }   {\widehat D}_{11} M_{11} A_{12},\ \  A'_{21} =  { \alpha_{k}  } A_{21} W_{11} {\widehat D}_{11}. \eqno(7)
 $$
 The middle matrix can be decomposed as follows
 
 $$  \left(\begin{array}{cc}  \alpha D_{11} & { 1 \over \alpha_{k}} A'_{12} \\  {1 \over \alpha_{k}} A'_{21} & A_{22} \end{array}\right) 
 = 
 \left(\begin{array}{cc}I & 0 \\  {1 \over {\alpha }} A'_{21}D^{+}_{11} & I \end{array}\right) 
 \left(\begin{array}{cc} 
 \alpha D_{11} & {\alpha \over \alpha_{k}} A''_{12} \\ { \alpha \over \alpha_{k}  } A''_{21} & {1 \over \alpha_{k}} A'_{22} \end{array}\right) 
\left(\begin{array}{cc}I & {1 \over {\alpha}} D^{+}_{11} A'_{12}  \\ 0 & I \end{array}\right) . \eqno(8) $$
 We denote 
 $$E_{11}=D_{11}^{\to 1},\ \ I_{11}= E_{11} E^T_{11},\ \ J_{11}= E^T_{11} E _{11}. $$
 
   To denote the generalized inverse matrix, we use the superscript plus. Note that for any matrix   $ (a_{i,j})\in {\bf S}_{wp}$ the generalized inverse matrix coincides with the pseudoinverse matrix:   
   $$(a_{i,j})^+=\{(b_{i,j}): b_{i,j}=0\ \hbox{ if }  a_{j,i}=0,\ b_{i,j}=a_{j,i}^{-1} \ \hbox{ if }  a_{j,i}\ne 0 \}.
   $$  
   A pseudoinverse matrix is obtained by transposing a given matrix and replacing all nonzero elements with inverse elements.

 As far as $\bar I_{11}  \widehat D_{11}=\bar D_{11}=\widehat D_{11}\bar J_{11}$, $D^+_{11}\bar I_{11}=0$ and $ D^+_{11}\widehat D_{11} = J_{11}$ we get    
 $$A''_{12}={ 1 \over (\alpha \alpha_k )}\bar I_{11}A'_{12}=   { 1 \over \alpha}  \bar D_{11} M_{11} A_{12},\  A''_{21}={ 1 \over (\alpha \alpha_k )}A'_{21} \bar J_{11}=  { 1 \over \alpha} A_{21} W_{11} \bar D_{11}. \eqno(9) $$

 For the lower right block we get
 $$ \alpha  \alpha_k^{2} A_{22}= (A'_{21}J_{11}+A''_{21})D^{+}_{11} A'_{12}+  A'_{21} D^{+}_{11}  A''_{12}+ \alpha \alpha_k A'_{22}$$
$$  A'_{22}=  { 1 \over \alpha\alpha_k } ({\alpha\alpha_k^{2}} A_{22} -  A'_{21}D^{+}_{11}  A'_{12}). \eqno(10)$$

Matrices $A''_{12}$ and $A''_{21}$   are matrices of surrounding minors with respect to minor $\alpha_k$. See the prove in Theorem 4.

Let  
 $$\left\{ \begin{array}{rcl} \alpha_k L_{21}D_{21}U_{21}= A''_{21} \\  L_{21}d_{21}M_{21}=    I \\  W_{21}d_{21} U_{21}=   I     \end{array}\right. \ \ \hbox{ and } \ \ 
 \left\{ \begin{array}{rcl} \alpha_k L_{12}D_{12}U_{12}= A''_{12} \\  L_{12}d_{12}M_{12}=    I \\  W_{12}d_{12} U_{12}=   I    \end{array}\right.$$ 
 be LDU decomposition of blocks $A''_{21}$ and  $A''_{12}$. 

 Let $A'$ be submatrix of $A$ which is fixed with non zero rows of  matrix $ diag(I_{11}, I_{12}, I_{21})$ and non zero columns of matrix  $ diag(J_{11}, J_{21}, J_{12})$.
 Submatrices $A''_{12}$, $A''_{21}$ and the submatrix, which corresponds to minor $\alpha_k$, do not have common nonzero rows and columns, so the sequence of nested non zero minors of submatrix $A'$ can be selected in different ways.
 
Let $rank(A''_{21})=l_1$ and $rank(A''_{12})=m_1$, then the rank of submatrix  $A'$ is equal $s=k+m_1+l_1$. Suppose that we have obtained the following sequences of nested minors: $det_{1},...,det_{k},  ,...,det_{l}$, $(l=l_1+k)$,  for the block $
(A_{11},  A_{21})^T$ and $det_{1},...,det_{k},det'_{k+1}...,det'_m$, $m= m_1+k $,  for the block $(A_{11}, A_{12})$.

 For the matrix $A'$ we can set the following sequences of nested minors:
$$det_{1},...,det_{k},...,det_{l},det_{l+1},,...,det_{s}, $$
with 
$$det_{l+i}=\lambda det'_{k+i },\ \ i=1,2,..,m_1,\ (\lambda= det_{l}/det_{k}),  $$
  in particular, $det_{s}=\lambda det'_{m} $. 
  
  We denote $\alpha_{l}=det_{l}, \ \alpha_{s}=det_{s}$, $J^{\lambda}_{12}= \lambda  J_{12}+\bar J_{12} $, $I^{\lambda}_{12}= \lambda  I_{12}+\bar I_{12}$,  
  $$ L^{\sim}_{12}=  L_{12}  I^{\lambda}_{12},\ \ U^{\sim}_{12}= J^{\lambda}_{12} U_{12}, \ \ D^{\sim}_{12}=\lambda^{-2} D _{12}, \eqno(11)$$
   $$  {\widehat D}^{\sim}_{12}=\lambda^{-1} I^{\lambda^{-1}}_{12} {\widehat D} _{12}, \ \ 
  M^{\sim}_{12}=\lambda M_{12},  \ \ W^{\sim}_{12}=\lambda W_{12}, \eqno(12)$$
  Then the last system can be written as follows:
 $$\left\{ \begin{array}{rcl} (\lambda \alpha_k) L^{\sim}_{12} D^{\sim}_{12}U^{\sim}_{12}= \lambda A''_{12} \\  L^{\sim}_{12}{\widehat D}^{\sim}_{12}M^{\sim}_{12}= I \\  W^{\sim}_{12}{\widehat D}^{\sim}_{12} U^{\sim}_{12}=   I    \end{array}\right..$$ 
So we can write the following matrix equation: 
  $$ \left(\begin{array}{cc} \alpha  D_{11} & { \alpha \over \alpha_{k} } A''_{12} \\ { \alpha \over \alpha_{k} } A''_{21} & { 1 \over \alpha_{k} } A'_{22} \end{array}\right)=\left(\begin{array}{cc}L^{\sim}_{12} & 0 \\ 0 & L_{21} \end{array}\right)   \left(\begin{array}{cc} 
  \alpha D_{11} & \alpha D^{\sim}_{12} \\ \alpha D_{21} & { 1 \over \alpha_{k} } A''_{22} \end{array}\right) 
  \left(\begin{array}{cc}U_{21} & 0 \\ 0 & U^{\sim}_{12} \end{array}\right), \eqno(13)$$
 where we denote
  $$A''_{22}=   {\widehat D}_{21}M_{21}A'_{22}  W^{\sim}_{12} {\widehat D}^{\sim}_{12}={\widehat D}_{21}M_{21}A'_{22}  W_{12}  I^{\lambda^{-1}}_{12} {\widehat D}_{12}
 \eqno(14)$$
 and use the following equation:  $I_{12}  I_{11}=0$,    $J_{11}  J_{21}=0$ and $L_{12} D_{11} U_{21}=D_{11}$. To check the last equation we can write each matrices as follows: $L_{12}=  L_{12} I_{12}+ \bar I_{12}$, $D_{11}=  I_{11}  D_{11} J_{11}$,   $U_{21}=   J_{21}  U_{21}+  \bar J_{21}$. 

 The middle matrix can be decomposed in two ways:
  $$ \left(\begin{array}{cc} \alpha D_{11} & \alpha  D^{\sim}_{12} \\ \alpha D_{21} & {1  \over  \alpha_k  } A''_{22} \end{array}\right)    =  \left(\begin{array}{cc}I & 0 \\   { 1 \over \alpha \alpha_k  } A''_{22}D^{\sim +}_{12} & I \end{array}\right)    \left(\begin{array}{cc} \alpha D_{11} & \alpha D^{\sim}_{12} \\ \alpha D_{21} & {\alpha \over \alpha_s}  A'''_{22} \end{array}\right)    \left(\begin{array}{cc}I &  { 1 \over  \alpha\alpha_k  } D^{+}_{21} A''_{22} \bar J_{12}  \\ 0 & I \end{array}\right) $$
 or
  $$   \left(\begin{array}{cc} \alpha D_{11} &  \alpha D^{\sim}_{12} \\ 
   \alpha  D_{21} & {1  \over  \alpha_k  } A''_{22} \end{array}\right)     
  =  
  \left(\begin{array}{cc}I & 0 \\   { 1  \over \alpha \alpha_k  } \bar I_{21}  A''_{22}D^{\sim +}_{12} & I 
  \end{array}\right) \left(\begin{array}{cc} 
  \alpha D_{11} &  \alpha  D^{\sim}_{12} \\  \alpha D_{21} & {\alpha  \over \alpha_s}  A'''_{22} 
  \end{array}\right)    \left(\begin{array}{cc}
  I &  { 1 \over \alpha\alpha_k } D^{+}_{21} A''_{22}  \\ 0 & I \end{array}\right). 
  \eqno(15)$$
  
    We use the following equations $D^{+}_{12} D_{11} = 0, \  D_{11} D^{+}_{21} = 0,$
and for the lower right block we get
 $$   {1  \over  \alpha_k^{2} } A''_{22}=   {1  \over  \alpha_k^{2} }  A''_{22} J_{12} +   {1  \over  \alpha_k^{2} }  I_{21}A''_{22}\bar J_{12}+ {\alpha \over \alpha_s}    A'''_{22}, $$
or
 $$ {1  \over  \alpha_k  } A''_{22}=  {1  \over  \alpha_k  } I_{21}  A''_{22} +  {1  \over  \alpha_k  } \bar  I_{21}A''_{22} J_{12}+ {\alpha  \over \alpha_s}   A'''_{22}.$$
 Both of these expressions give the same value for $ A'''_{22}$:
 
  $$   A'''_{22}=   {\alpha_s  \over  \alpha\alpha_k  } \bar I_{21}A''_{22}\bar J_{12}= {1 \over  \alpha_k^{2} \alpha  } \bar D_{21}M_{21} A'_{22} W_{12} \bar D_{12}. \eqno(16)$$
 Further we will use the second decomposition.
 
 The  matrix $A'''_{22}$ is the matrix of surrounding minors of $A$ with respect to minor $\alpha_s$. See prove in Theorem 4.

 Let  
 $$ \left\{ \begin{array}{rcl} \alpha_s L_{22} D_{22} U_{22}=  A'''_{22} \\  L_{22}d_{22} M_{22}=  I \\  W_{22} d_{22} U_{22} =   I    \end{array}\right. $$
 be a decomposition of the matrix $ A'''_{22}$, then we can write the equality 
 $$    \left(\begin{array}{cc}  D_{11} &   D^{\sim}_{12} \\   D_{21} & { 1 \over \alpha_s}   A'''_{22} \end{array}\right)   = \left(\begin{array}{cc}I & 0 \\ 0 & L_{22} \end{array}\right)   \left(\begin{array}{cc}   D_{11} &    D^{\sim}_{12} \\ D_{21} &  D_{22} \end{array}\right)    \left(\begin{array}{cc}I & 0 \\ 0 & U_{12} \end{array}\right). \eqno(17)$$

  To prove this equality  we can write   matrices $L_{22}$, $U_{22}$, $D_{12}$, $D_{12}$ as follows 
 $ L_{22}=   L_{22} I_{22}+  \bar I_{22},\  U_{22}= J_{22} U_{22} + \bar   J_{22},$  
$ D_{21}=  I_{21}  D_{21}  J_{21},\ D_{12}=   I_{12}  D_{12}  J_{12},$ 
 and check the equations 
 $$L_{22} D_{21}=D_{21} \hbox{  and  }   D^{\sim}_{12}U_{22}=D^{\sim}_{12}.$$
As a result of the  sequence (6), (8), (13), (15),(17) of decompositions we obtain the LDU-decomposition of the matrix A in the form
  $$    \left(\begin{array}{cc}A_{11} & A_{12} \\ A_{21} & A_{22} \end{array}\right)  = 
  \alpha L  D  U, \ \ D=
  \left(\begin{array}{cc} D_{11} & D^{\sim}_{12} \\ D_{21} & D_{22} \end{array}\right)=
    \left(\begin{array}{cc} D_{11} & \lambda^{-2} D_{12} \\ D_{21} & D_{22} \end{array}\right), 
  \eqno(18)$$ 
 with $D^{\sim}_{12}=\lambda^{-2}  D_{12}$  and such matrices $L$ and $U$:
 $$L=  \left(\begin{array}{cc}L_{11} & 0 \\ 0 & I \end{array}\right)   \left(\begin{array}{cc}I & 0 \\ \alpha_k^{-1} A'_{21}D^{+}_{11} & I \end{array}\right)    \left(\begin{array}{cc}L^{\sim}_{12}  & 0 \\ 0 & L_{21} \end{array}\right)   \left(\begin{array}{cc}I & 0 \\  \alpha^{-1} \alpha_k^{-1} \bar I_{21} A''_{22}D^{{\sim +}}_{12} & I \end{array}\right)    \left(\begin{array}{cc}I &  0 \\ 0 &  L_{22} \end{array}\right) ,$$ 
 $$U= \left(\begin{array}{cc}I &  0 \\ 0 &  U_{22} \end{array}\right)    \left(\begin{array}{cc}I &  \alpha^{-1} \alpha_k^{-1} D^{+}_{21}A''_{22} \\ 0 & I \end{array}\right)   \left(\begin{array}{cc}U_{21} & 0 \\ 0 & U^{\sim}_{12} \end{array}\right)       \left(\begin{array}{cc}I & \alpha_k^{-1} D^{+}_{11} A'_{12} \\ 0 & I \end{array}\right)   \left(\begin{array}{cc}U_{11} & 0 \\ 0 & I \end{array}\right). $$
 After multiplying the matrices on the right side, we get 
 $$ L= \left(\begin{array}{cc}   L_{1}  & 0 \\ L_{3} & L_{4} \end{array}\right)=\left(\begin{array}{cc}   L_{11}L^{\sim}_{12} & 0 \\ L_{3} & L_{21}L_{22} \end{array}\right), \ \   U =\left(\begin{array}{cc} U_{1} & U_{2} \\   0 & U_{4} \end{array}\right)=  \left(\begin{array}{cc}U_{21}U_{11} & U_{2} \\   0 & U_{22}U^{\sim}_{12} \end{array}\right), \eqno(19)$$
with
 $$ U_{2}= \alpha_k^{-1} U_{21} D^{+}_{11} A'_{12}+ \alpha^{-1}\alpha_k^{-1}  D^{+}_{21}A''_{22} U^{\sim}_{12}  = $$
 $$=\alpha_k ^{-1} J_{11} M_{11} A_{12}+  \alpha^{-1}\alpha_l^{-1} J_{21} M_{21}A'_{22},  \eqno(20)$$
  $$L_{3}=\alpha_k^{-1} A'_{21} D^{+}_{11}L^{\sim}_{12}+  \alpha^{-1} \alpha_k^{-1}  L_{21} \bar I_{21} A''_{22}D^{\sim +}_{12} =$$
  $$=\alpha_k^{-1} A_{21}W_{11} I_{11} + \alpha^{-1} \alpha_m^{-1} \alpha_k^{-1} \bar D_{21} M_{21} A'_{22}  W_{12}   I_{12} \eqno(21)$$
   $$ {\widehat D}=\alpha_r^{-1}(\alpha D +\bar D) \eqno(22)$$
 $$M={\widehat D}^{-1}L^{-1},= {\widehat D}^{-1}  \left(\begin{array}{cc}   L_{1}^{-1}  & 0 \\ -L_{4}^{-1}L_{3}L_{1}^{-1} & L_{4}^{-1} \end{array}\right)= 
 $$
 $$ {\widehat D}^{-1}
  \left(\begin{array}{cc} \lambda^{-1}{\widehat D}_{12} M_{12} {\widehat D}_{11} M_{11}  & 0 \\ 
 -{\widehat D}_{22}M_{22} {\widehat D}_{21} M_{21}L_{3} \lambda^{-1}
 {\widehat D}_{12} M_{12} {\widehat D}_{11} M_{11} &
 {\widehat D}_{22} M_{22} {\widehat D}_{21} M_{21} 
 \end{array}\right),  \eqno(23)
$$
  $$W=U^{-1}{\widehat D}^{-1}=  \left(\begin{array}{cc}   U_{1}^{-1}  &  -U_{1}^{-1}U_{2} U_{4}^{-1} \\ 0 & U_{4}^{-1} \end{array}\right) 
  {\widehat D}^{-1}=
  $$
  $$
  \left(\begin{array}{cc}  W_{11} \widehat D_{11} W_{21} \widehat D_{21}  &  -W_{11} \widehat D_{11} W_{21}U_{2} W_{12} \widehat D_{12}J^{\lambda^{-1}}_{12} W_{22} \widehat D_{22} \\
  0 &  W_{12} \widehat D_{12}J^{\lambda^{-1}}_{12} W_{22} \widehat D_{22} \end{array}\right) {\widehat D}^{-1}. \eqno(24)$$

 In the   expressions (18) - (24),  the $ LDU $ decomposition of the matrix $ A $ are given and the matrices $ W $ and $ M $  that satisfy the conditions $ LdM = I $ and $ WdU = I $ are obtained.
 
 We proved the correctness of the following recursive algorithm.
\section{ Algorithm of LDU-decomposition  }  
      $$(L, D, U, M, {\widehat D}, W, \alpha_r)=\, \mathbf{LDU}(A, \alpha).$$

      \noindent
 {\bf (1)}   If ($A=0$)  then 
 
 \noindent $\{$  $\alpha_r = \alpha;  $ \ \ 
 $  D = I = J= 0;$
 
 $ L =U = \bar I = \bar J= \bar D  = I; $ \ \  
               $ M = W =\widehat  D = \alpha I;$ $\}$
              
 
\noindent 
 {\bf (2)} If ($n=1$ $\&$ $A=[a]$  $\&$ $a\ne 0$)  then 
 
 \noindent $\{$
             $ \alpha_r =a;\ L = U = M = W = [a];\
                D = [(\alpha*a)^{-1}];\ \widehat  D=[a^{-2}];$
                
                $J=I=[1];\
                \bar I= \bar J = \bar D = [0];\  \}$  
  
\noindent
 {\bf (3)} If ($n\ge 2$ $\&$ $A \ne 0$) then  
 
 \noindent $\{$
 $A= \left(\begin{array}{cc}A_{11} & A_{12} \\ A_{21} & A_{22} \end{array}\right) $. 
 
 (3.1)  $$(L_{11}, D_{11}, U_{11}, M_{11}, d_{11}, W_{11}, \alpha_k )=\, \mathbf{LDU}(A_{11}, \alpha),$$ 

      $A_{12}^0= M_{11}*A_{12};$ 

     $ A_{12}^1= a_k* \widehat  D_{11} * A_{12}^0 ;$
      
      $A_{12}^2= \bar D_{11} * A_{12}^0 / \alpha ;$
      
      $A_{21}^0=A_{21}*W_{11};$  
      
      $A_{21}^1 =a_k * A_{21}^0 *\widehat  D_{11} ;$
      
      $A_{21}^2= A_{21}^0 * \bar D_{11} / \alpha ;$

       (3.2)  $$(L_{21}, D_{21}, U_{21}, M_{21}, d_{21}, W_{21}, \alpha_l )=\, \mathbf{LDU}(A_{21}^2, a_k),$$
       
        (3.3)  $$(L_{12}, D_{12}, U_{12}, M_{12}, d_{12}, W_{12}, \alpha_m )=\, \mathbf{LDU}(A_{12}^2, a_k),$$     
 
        $\lambda=  {a_l \over a_k}  ; a_s=\lambda*a_m ;$  
 
         $A_{22}^0= A_{21}^1*D_{11}^{+}*A_{12}^1;$
         
          $A_{22}^1= ( \alpha a_k^2*A_{22} - A_{22}^0 )/(\alpha a_k) ;$
          
         $ A_{22}^2=\bar D_{21}*M_{21}*A_{22}^1*W_{12}* \bar  D_{12} ;$
         
          $A_{22}^3=A_{22}^2/(a_k^2 \alpha);$      

   (3.4)  $$(L_{22}, D_{22}, U_{22}, M_{22}, d_{22}, W_{22}, \alpha_r )=\, \mathbf{LDU}(A_{22}^3, a_s),$$ 
 
                $J_{12}^{\lambda}=  \lambda* J_{12}  + \bar J_{12} ;\ \ 
                I_{12}^{\lambda}= \lambda I_{12}   + \bar I_{12} ;$  
                
               $  L_{12}^{\sim} = L_{12}*I_{12}^{\lambda} ;\ \ 
                 U_{12}^{\sim} = J_{12}^{\lambda}*U_{12} ;$        
   
        $U_{2}=J_{11}*M_{11}*A_{12}/a_k + J_{21}*M_{21} *A_{22}^1 /(a_l*\alpha);$
        
        $ L_{3}= A_{21}*W_{11}*I_{11}/a_k +\bar D_{21}*M_{21} *A_{22}^1*W_{12}*I_{12}/(a_m* a_k* \alpha);$

          $$
          L= \left(\begin{array}{cc}   L_{11}  L^{\sim} _{12} & 0 \\ L_{3} & L_{21}L_{22} \end{array} \right),  \ \ 
 D=\left(\begin{array}{cc}   D_{11} &   \lambda^{-2}  D_{12} \\    D_{21} &   D_{22} \end{array}\right) , \ \ 
 U =  \left(\begin{array}{cc}U_{21}U_{11} & U_{2} \\   0 & U_{22}   U^{\sim}_{12} \end{array}\right), 
 $$
     $$  \widehat D= \alpha (\alpha_r)^{-1}D+ (\alpha_r)^{-1}\bar D, $$ 
       $$M=  {\widehat D}^{-1}
  \left(\begin{array}{cc} I^{\lambda^{-1}}_{12}{\widehat D}_{12} M_{12} {\widehat D}_{11} M_{11}  & 0 \\ 
 -{\widehat D}_{22}M_{22} {\widehat D}_{21} M_{21}L_{3} I^{\lambda^{-1}}_{12}
 {\widehat D}_{12} M_{12} {\widehat D}_{11} M_{11} &
 {\widehat D}_{22} M_{22} {\widehat D}_{21} M_{21} 
 \end{array}\right), 
$$
  $$W= \left(\begin{array}{cc}  W_{11} \widehat D_{11} W_{21} \widehat D_{21}  &  -W_{11} \widehat D_{11} W_{21}U_{2} W_{12} \widehat D_{12}J^{\lambda^{-1}}_{12} W_{22} \widehat D_{22} \\
  0 &  W_{12} \widehat D_{12}J^{\lambda^{-1}}_{12} W_{22} \widehat D_{22} \end{array}\right) {\widehat D}^{-1}. $$
      
   \noindent $\}$ 
        

  %

  \subsection{Auxiliary Theorems}

The following  statements prove the factorization algorithm.

 \begin{theorem}  
 Let $ {\cal M}_k $ and $ {\cal M}_s $ be corner blocks of size $ k \times k $ and $ s \times s $,  ($s>k$, $ s = k + t $) of the matrix $ \cal M $, their determinants $ det_k $ and $ det_s $ not equal to zero. Let the matrix $ A $ be formed by the  surrounding minors of the block $ {\cal M}_k $ and let it be divided into blocks 
     $ A= \left(\begin{array}{cc}A^1  & A^2 \\ A^3 & A^4 \end{array}\right)$, wherein
    $ A^1$ is a square block of size $ t \times t $ and $ A^{1*} $ is its adjoint matrix,
    then elements of matrix 
     $$ {1 \over  det_k }(det_s A^4 - {1 \over  det_k^{t-1} } A^3 A^{1*} A^2) \eqno(25)$$
     are the minors of the matrix $ \cal M $ that surround the block $ {\cal M}_s $.
 \end{theorem}
 {\bf Proof.} 
 Proof can be found in 
  (\cite{6}, Theorem 2) or in (\cite{7} pp. 23-25).  In \cite {7}, this theorem is called the `` Determinant Identity of Descent ''.
  %
  
 You can see that Theorem 3 generalizes Theorem 2 if we assume that the block $ A_k $ can have size 0 and the determinant of such an empty block is 1. And Theorem 2 is a special case of Theorem 3 if we consider each element of the original matrix as a surrounding  minor for an empty block. 
  
 \begin{theorem}  
 Let the matrix $ A $ be formed by the  surrounding minors of the upper left corner block $ \alpha $ of the matrix $\cal M$. 
 Let matrix $A$ be divided into blocks (4) and all equalities of system (5) are true, 
 $rank(A_{11})=k$ and $\alpha_k$ is the largest non zero minor of $A_{11}$.
Then matrices $A''_{12}$ and $A''_{21}$ (9) are matrices of surrounding minors with respect to minor $\alpha_k$,  $A'''_{22}$ (16) is the matrix of surrounding minors with respect to
minor  $\alpha_s$.
 \end{theorem}

{\bf Proof.}

 To simplify writing the proof, we consider the case when the non-zero block $D^1$ of matrix
 $D_{11}$ be in the upper left corner of $D_{11}$ and we denote by $A^1$ a non-degenerate block 
 of size $ k \times k $ in upper left corner of the matrix $ A_{11} $.

 We can write the LDU decomposition of matrix $A_{11}$ and the  equalities  $W_{11}d_{11}U_{11}=I$ and $L_{11}d_{11}M_{11}=I$ in such block shape:
  $$ A_{11} = \left(\begin{array}{cc}A ^ 1_{11}  &  A ^2_{11} \\ A ^3_{11} &  A ^4_{11}  \end{array}\right) 
  =\alpha \left(\begin{array}{cc}L^ 1_{11}  &  0 \\ L^3_{11} &  I  \end{array}\right) \left(\begin{array}{cc}D^1_{11}    &  0 \\ 0 &  0  \end{array}\right)
  \left(\begin{array} {cc}U^1_{11}  &  U^2_{11} \\ 0 &  I \end{array}\right),
  $$
   $$ \left(\begin{array}{cc}W^1_{11}  &  W^2_{11} \\ 0 &   \alpha_k I  \end{array}\right) \left(\begin{array}{cc} \alpha_k^{-1} \alpha D^1_{11} & 0 \\ 0 &  \alpha_k^{-1} I \end{array}\right) \left(\begin{array} {cc}U^1_{11}  &  U^2_{11} \\ 0 &  I \end{array}\right)=  \left(\begin{array}{cc}   I  &  0 \\ 0 &   I  \end{array}\right),
   $$
     $$ \left(\begin{array}{cc}L^1_{11}  &  0 \\ L^ 2_{11} & I \end{array}\right) \left(\begin{array}{cc}  \alpha_k^{-1} \alpha D^1_{11} & 0 \\ 0 & \alpha_k^{-1} I \end{array}\right) \left(\begin{array} {cc}M^1_{11}  &  0 \\ M^2_{11} &  \alpha_k I \end{array}\right) =  \left(\begin{array}{cc} I & 0 \\ 0 & I \end{array}\right).$$ 

    The determinant of the $ {\cal M} $ submatrix, which is defined by all rows and 
  columns of the minors $\alpha$ and the block $A^1_{11}$ is equals $\alpha_k$. 
 Due to the Sylvester determinant identity (see \cite{1}) we can write the equality:
 $$
  det(A^1_{11})=\alpha_k \alpha^{k-1}. 
 $$ 
     
   From the first equality and Sylvester determinant identity we get 
   $$
   \alpha L^1_{11} D^1_{11} U^1_{11}=A^1_{11}  , \ \ \alpha L^1_{11} D^1_{11} U^2_{11}=A^2_{11} , 
   $$ 
   $$
   \alpha_k \alpha^{k-2} (U^1_{11})^{-1} (D^1_{11} )^{-1} (L^1_{11})^{-1}=A^{1*}_{11} ,  \ \    \alpha_k  \alpha^{k-2} (U^1_{11})^{-1}(D^1_{11} )^{-1} = A^{1*}_{11} L^1_{11}.  
   $$
   The matrix $A^{1*}_{11} $ is the adjoin matrix for $A^{1}_{11} $.    
   From the second equality we get 
   $$
   \alpha W^1_{11} D^1_{11} U^1_{11}= \alpha_r I,\ \ \alpha W^1_{11}= \alpha_k (U^1_{11})^{-1}(D^1_{11} )^{-1},\ \   W^2_{11}=- \alpha W^1_{11} D^1_{11} U^2_{11}.$$
   
   The consequence is the expressions for the blocks $W^1_{11}$ and $W^2_{11}$:
   $$
   W^1_{11}= \alpha^{1-k}  A^{1*}_{11} L^1_{11}, \ \  W^2_{11}=- \alpha^{1-k}   A^{1*}_{11} A^2_{11},\ \  W_{11}=\left(\begin{array}{cc} W^1_{11}  &  W^2_{11} \\ 0 &  \alpha_k I  \end{array}\right) .  \eqno(26)
   $$
   From the third equality we obtain the expressions $ \alpha L^1_{11} D^1_{11} M^1_{11}= \alpha_k I$,  $M^2_{11}=- \alpha L^2_{11} D^1_{11} M^1_{11}$, so
$$ M^1_{11}= \alpha^{1-k}  U^1_{11} A^{1*}_{11} , \ \ M^3_{11}=- \alpha^{1-k}  A^3_{11}  A^{1*}_{11},\ \ M_{11}= \left(\begin{array}{cc}  M^1_{11}  & 0 \\ -   M^3_{11}   &  \alpha_k I  \end{array}\right).  \eqno(27)$$

   Let the matrix $A_{21}$ be divided into two blocks $A_{21}=({ A^{\bf 1}_{21}, A^{\bf 2}_{21}})=\left(\begin{array}{cc}A^a_{21}   &  A^b_{21} \\ A^c_{21} &  A^d_{21}  \end{array}\right)$, then matrix $A''_{21}= { 1 \over \alpha} A_{21}W_{11}\bar D_{11}$ can be written as follows: 
   $$
   A''_{21}= { 1 \over \alpha} ({ A^{\bf 1}_{21}, A^{\bf 2}_{21}}) \left(\begin{array}{cc} \alpha^{1-k} A ^{1*} L ^1  &  -  \alpha^{1-k} A ^{1*} A ^2 \\ 0 &  \alpha_k I  \end{array}\right)   \left(\begin{array}{cc}0   &  0 \\ 0 &  I  \end{array}\right) =
   $$
   $$
(0, \ \     
 { 1 \over \alpha}  ( \alpha_k { A^{\bf 2}_{21}-{ 1 \over \alpha^{k-1}} A^{\bf 1}_{21}} A^{1*}_{11} A^2_{11} ))  .$$
According to Theorem 3, $A''_{21}$ is a matrix of surrounding minors  with respect to the block $A^1_{11}$. 
 
 Let the matrix $A_{12}$ be divided into two blocks $A_{12}= \left(\begin{array}{c}A^{\bf 1}_{12} \\ A^{\bf 2}_{12} \end{array}\right)=\left(\begin{array}{cc}A^a_{12} & A^b_{12} \\ A^c_{12} &  A^d_{12} \end{array}\right)$, then matrix $A''_{12}={1 \over  \alpha} \bar D_{11}M_{11}A_{12}$ can be written as follows:

    $$A''_{12}= {  1 \over \alpha} \left(\begin{array}{cc}0 & 0 \\ 0 & I \end{array}\right) 
    \left(\begin{array}{cc}\alpha^{1-k} U_{11}^1 A_{11}^{1*} & 0 \\ -\alpha^{1-k} A_{11}^3  A_{11}^{1*} &  \alpha_k I \end{array}\right)
     \left(\begin{array}{c}{  A^{\bf 1}_{12}} \\{ A^{\bf 2}_{12}} \end{array}\right)=
     $$
     $$
     \left(\begin{array}{c}0 \\  { 1 \over \alpha}   (  \alpha_k { A^{\bf 2}_{12}}- { 1 \over \alpha^{k-1}} A^3_{11}A^{1*}_{11} { A^{\bf 1}_{12}}) \end{array}\right).$$
 
 According to Theorem 3, $A''_{12}$ is a matrix of surrounding minors with respect to the block $A^1_{11}$. 

 Matrices $ W_{11}$,  $D_{11}$ and $M_{11}$  have such block shape:
  $$ W_{11}  D_{11} M_{11}=   \left(\begin{array}{cc}W^ 1  &  W^2 \\ 0 &  \alpha_k I  \end{array}\right) \left(\begin{array}{cc}  D^1_{11}  &  0 \\ 0 & 0   \end{array}\right) 
  \left(\begin{array}{cc}M^ 1  &  0 \\ M^3 & \alpha_k I \end{array}\right)=
  \left(\begin{array}{cc} W^1   D^1_{11} M^1   &  0 \\ 0 & 0 \end{array}\right).$$
$$ W^1   D^1_{11} M^1 = \alpha^{2-2k} A^{1*}_{11}L^1  D^1_{11} U^1 A^{1*}_{11} =  \alpha_k  \alpha^{-k} A^{1*}_{11}.   $$
 By definition (10) and expression (7) we have 
 
 $$  A'_{22}=  { 1 \over \alpha\alpha_k } ({\alpha\alpha_k^{2}} A_{22} -  A'_{21}D^{+}_{11}  A'_{12})={ 1 \over \alpha  } ({\alpha\alpha_k } A_{22} -    A_{21} W_{11} {\widehat D}_{11} J_{11}{\alpha \over \alpha_k}  M_{11} A_{12})=
 $$ 
 $$
  { \alpha_k } A_{22} -   {\alpha \over \alpha_k^2} A_{21} (W_{11} {D}_{11}  M_{11}) A_{12}=
 $$
$$
\alpha_k  A_{22} - \alpha^{1-k}  A_{21} 
\left(\begin{array}{cc} 
A^{1*}_{11}   &  0 \\ 0 & 0 \end{array}\right)  A_{12}=
\left(\begin{array}{cc} A'^1_{22}  &  A'^2_{22} \\ A'^3_{22} &  A'^4_{22} 
\end{array}\right). 
$$
According to Theorem 3, $\alpha^{-1} A'_{22}$ is a matrix of surrounding minors with respect to the minor $\alpha_k$.  
 
Let us denote by $A''^{i}_{21}$ and $A''^{i}_{12}$ ($i=1..4$) the blocks of the matrices $A''_{21}$ and  $A''_{12}$,
correspondingly, and denote by  $A''^{I}_{22}$ and $A'^{II}_{22}$   the upper and lower blocks. of the matrix 
$\alpha^{-1} A'_{22}$:  $\alpha^{-1} A'_{22}=[A''^{I}_{22}, A'^{II}_{22}]^T$.
Similarly to expressions (26) and (27), we obtain the following expressions:
 $$
 \bar D_{21} M_{21}=\left(\begin{array}{cc}0 & 0 \\ -\alpha_k^{1-l_1} A''^{3}_{21} A''^{1*}_{21} & \alpha_l I  \end{array}\right),
 \ \ W _{12}\bar D_{12}= 
 \left(\begin{array}{cc}0  & - \alpha_k^{1-m_1} A''^{1*}_{12}A''^2_{12} \\ 0 & \alpha_m I  \end{array}\right) .
 $$
According to Theorem 3,
$$A'^{L}_{22}=  \alpha_k^{-1} \bar D_{21} M_{21} (\alpha^{-1} A'_{22})  = \alpha_k^{-1} \left(\begin{array}{cc}0 & 0 \\ -\alpha_k^{1-l_1} A''^{3}_{21} A''^{1*}_{21} & \alpha_l I  \end{array}\right) \left(\begin{array}{c}A'^{I}_{22}  \\  A'^{II}_{22}  \end{array}\right)=
$$
$$
\left(\begin{array}{c}0  \\ {1 \over \alpha_k}( \alpha_l A'^{I}_{22} -{1 \over \alpha_k^{ l_1-1}} A''^{3}_{21} A''^{1*}_{21} A'^{II}_{22})  \end{array}\right) 
$$
is a matrix of surrounding minors with respect to the minor $\alpha_l$.  
 
 Finally, let us turn to the matrix (16):  
 $$   A'''_{22}=   {1 \over  \alpha_k }({1 \over  \alpha_k  } \bar D_{21}M_{21} ( {1 \over   \alpha  }A'_{22})) W_{12} \bar D_{12}={1 \over  \alpha_k } A'^{L}_{22}W_{12} \bar D_{12}= 
 $$
$$
{1 \over  \alpha_k  }\left(\begin{array}{cc } A'^{L_1}_{22}  & A'^{L_2}_{22}  \end{array}\right) 
 \left(\begin{array}{cc}0  & - \alpha_k^{1-m_1} A''^{1*}_{12}A''^2_{12} \\
 0 & \alpha_m I  \end{array}\right)= 
 $$
 $$
 \left(\begin{array}{cc}0  & {1 \over  \alpha_k  }(\alpha_m A'^{L_2}_{22} -{1 \over \alpha_k^{ m_1-1}}  A'^{L_1}_{22} A''^{1*}_{12}A''^2_{12})   \end{array}\right)=
 $$
  $$
 \left(\begin{array}{cc}0  & {1 \over  \alpha_l  }(\alpha_s A'^{L_2}_{22} -{1 \over \alpha_l^{ m_1-1}}  A'^{L_1}_{22} A''^{1\sim *}_{12}A''^{2\sim}_{12})   \end{array}\right).
 $$
 
 Here we introduced notations $A'^{L_1}_{22}$  and $A'^{L_2}_{22}$  for the left and right blocks of matrix $A'^{L}_{22}$, used definitions (11), (12) and Sylvester determinant identity. According to Theorem 3, $A'''_{22}$ is a matrix of surrounding minors with respect to the minor $\alpha_s$.
 
 \section*{Example}
 
We give below an example of a LDU-decomposition of a matrix in the form of three identities
$A= L D U$, $ L \widehat D M =  {\bf I }$, $W \widehat D U =   {\bf I }$:
$$ \left(\begin{array}{cccc} 0&2& 3&0  \\ 0&0& 0&-3 \\ 5&3& 2&1  \\ 0&-1&0&0 \end{array}\right)= 
 \left(
\begin{array}{cccc}2&0&0&0\\0&-30&0&0\\3&0&10&0\\-1&0&0&-45\end{array}
 \right)   
\left(\begin{array}{cccc} 0&      1 \over 2 &0&       0        \\ 0&     0&    0&   -1 \over 300  \\ 1 \over 20 &0&    0&       0        \\ 0&     0& 1 \over 1350 &0       \end{array}\right) 
\left(\begin{array}{cccc} 10&0&-5& 2   \\ 0& 2&3&  0   \\ 0& 0&-45&0   \\ 0& 0&0&  -30\end{array}\right),$$
$$ \left(
\begin{array}{cccc}2&0&0&0\\0&-30&0&0\\3&0&10&0\\-1&0&0&-45\end{array}
 \right) 
\left(\begin{array}{cccc} 0& -1 \over 90 &0&         0         \\ 0&       0&      0&  1 \over 13500  \\  -1 \over 900 &0&      0&         0         \\ 0&       0&       -1 \over 60750 &0        \end{array}\right)
\left(\begin{array}{cccc} 135&0&   -90&0    \\ -45&0&   0&  0    \\ 675&0&   0&  1350 \\ 0&  -450&0&  0   \end{array}\right)=\bf I, $$
$$ 
\left(\begin{array}{cccc} 0&  90&  -90&675   \\ -45&0&   0&  -2025 \\ 0&  0&   0&  1350  \\ 0&  -450&0&  0    \end{array}\right) 
\left(\begin{array}{cccc} 0& -1 \over 90 &0&         0         \\ 0&       0&      0&  1 \over 13500  \\  -1 \over 900 &0&      0&         0         \\ 0&       0&       -1 \over 60750 &0 \end{array}\right) 
\left(\begin{array}{cccc} 10&0&-5& 2   \\ 0& 2&3&  0   \\ 0& 0&-45&0   \\ 0& 0&0&  -30\end{array}\right)=\bf I.
$$
 
 The products of matrices $\widehat D  M$ and $W \widehat D$ can be reduced to a triangular form by inserting the product of the permutation matrix and the inverse permutation matrix between the factors:
 
$$\widehat D  M=\left(\begin{array}{cccc}  
  -1 \over 90 &0& 0 &     0                 \\
   0    &1 \over 13500 & 0 &      0          \\
   0          & 0 &-1 \over 900 &      0                 \\
   0    & 0 & 0 &-1 \over 60750       
\end{array}\right)
\left(\begin{array} {cccc}
-45   &    0 & 0 &  0 \\
0     & -450 & 0 &  0 \\
135   &    0 &-90&  0 \\
675   &    0 & 0 &  1350 
\end{array}\right) $$
 
 $$W \widehat D=
 \left(\begin{array}{cccc}
-90 & 0  &     675  &  90\\
 0&-45&      -2025 &   0\\
 0&0  &       1350  &    0\\
 0&0    &   0    &  -450
 \end{array}\right) 
\left(\begin{array}{cccc}
-1 \over 900 &0&      0&         0         \\
0& -1 \over 90 &0&         0         \\
0&       0&       -1 \over 60750 &0 \\
0&       0&      0&  1 \over 13500  
\end{array}\right).$$
 
  \section*{Conclusion}
 
 A dichotomous $LDU$ factorization algorithm was proposed. 
It is applied to matrices in which the size is some power of 2. Such an algorithm is well parallelized and efficient for a supercomputer with distributed memory due to the presence of a coarse-grained block structure. If you want to find the decomposition of an arbitrary rectangular matrix, you must first arrange it arbitrarily inside a square matrix of a suitable size, perform the decomposition, and in the resulting factors, you need to discard the extra zero parts of the matrices.
 
 As with all previous recursive algorithms, its complexity (up to a constant) is equal to the complexity ($n^\omega$) of matrix multiplication. Like other $LU$ algorithms, it gives a gain of $r/n$ times when applied to matrices of small rank r. But it is also efficient for full rank sparse matrices. An example of this type of matrices that is important in applications is considered in the work \cite{3}.

Since the decomposition of the upper right block and the lower left block will be performed simultaneously, it is desirable that these two blocks have more nonzero elements than the other two blocks. If there is an almost-diagonal, tridiagonal or ribbon matrix, then it must first be multiplied by a permutation matrix so that the main diagonal is located in these two blocks.
 
It should be noted that this algorithm is a generalization of the $LEU$  algorithm \cite{2} to the case of a commutative domain. Therefore, it can be looked at as another proof of the $LEU$ algorithm. We specifically emphasized the non-unity of the decomposition due to the fact that Eq. (15) can be applied in either of two versions. The complexity of the proof of the $LEU$ algorithm was the reason that some authors proposed their own proofs, in which they stated the uniqueness of the decomposition and even came up with a new name for the matrix $E$. Matrix $E$ is called the `` Bruhat Permutation Matrix '', since it first appeared in the works of Bruhat ( see \cite {7A}, \cite {7B}).

This LDU factorization algorithm is another step in creating a common library of block-recursive linear algebra algorithms. The first in this area were A.A. Karatsuba \cite{8A} and W. Strassen \cite{8}. The understanding of the importance of recursive algorithms for supercomputer computing came only in recent decades and led to a program for creating decentralized dynamic control technology for the supercomputer's computing process (see \cite{2009}, \cite{2018I}, \cite{2018}). Other examples of recursive algorithms in the commutative domain, such as computing the inverse and adjoint matrices, the kernels of a linear operator, can be found in \cite {2006}, \cite {2008}, \cite {2018I}. It is expected that new recursive algorithms should appear in the class of problems of orthogonal matrix factorization.
 
 It is important to note that this algorithm does not accumulate errors and all computations take place in the commutative domain. The application of the Chinese remainder theorem and the transition to finite fields provides a way to reduce the total number of operations and very efficient parallelization on a supercomputer. Note that it is not necessary to search for original images for the elements of matrix $D$ by their images in finite fields, since they are easily found by the diagonal elements of matrices $L$ and $U$. Thus, the upper bound for the maximum minor of matrix $A$ can be used to estimate the largest element which appear at the end of the computational process.

The discussed algorithms are used in the cloud computer algebra Math Partner \cite {2017}. You can fined this system at: mathpar.ukma.edu.ua.
 

  \end{document}